\documentclass[a4paper,11pt]{article}
\pdfoutput=1 % to allow pdflatex compilation in JCAP
\usepackage{jcappub}
\usepackage{amsmath}
\usepackage{graphicx}
\usepackage{comment}
\usepackage{latexsym}
\usepackage{xspace}
\usepackage{lscape}
\usepackage{color}
\usepackage{hyperref}
\usepackage{bm}
\usepackage{relsize}
\usepackage{tabularx}
\usepackage{multirow}
\usepackage{amssymb}
\usepackage[table]{xcolor}
\usepackage{lipsum} 

\newcommand{\ee}[1]{\begin{equation}#1\end{equation}}
\newcommand{\ea}[1]{\begin{align}#1\end{align}}

\newcommand{\TM}[1]{\textbf{\color{blue}Tommi: #1}}

\providecommand{\f}[2]{\frac{{#1}}{{#2}}}

\def\beq{\begin{equation}}
\def\eeq{\end{equation}}
\def\baq{\begin{eqnarray}}
\def\eaq{\end{eqnarray}}

\makeatletter
\gdef\@fpheader{}
\g@addto@macro\bfseries{\boldmath}
\makeatother

\title{Primordial dark matter from curvature induced symmetry breaking}

\author[a,b]{Laura Laulumaa,}
\author[c,b]{Tommi Markkanen,}
\author[a,b]{and Sami Nurmi}

\affiliation[a]{Department of Physics, University of Jyv\"{a}skyl\"{a}, P.O. Box 35, FI-40014 University of
Jyv\"{a}skyl\"{a}, Finland}
\affiliation[b]{Helsinki Institute of Physics (HIP)
P.O. Box 64, 00014 University of Helsinki,  Finland}
\affiliation[c]{Laboratory of High Energy and Computational Physics, National Institute of Chemical Physics and
Biophysics, R\"avala pst. 10, Tallinn, 10143, Estonia}

\emailAdd{laura.e.laulumaa@student.jyu.fi}
\emailAdd{tommi.markkanen@kbfi.ee}
\emailAdd{sami.t.nurmi@jyu.fi}

\abstract{
We demonstrate that adiabatic dark matter can be generated by gravity induced symmetry breaking during inflation. We study a $Z_2$ symmetric scalar singlet that couples to other fields only through gravity and for which the symmetry is broken by the spacetime curvature during inflation when the non-minimal coupling $\xi$ is negative.  We find that the symmetry breaking leads to the formation of adiabatic dark matter with the observed abundance for the singlet mass $m\sim{\rm MeV}$ and $|\xi|\sim 1$.  
}


\begin{document}

%\arxivnumber{16XX.XXXXX}

\maketitle

\section{Introduction}
\label{sec:intro}
The conventional approach to dark matter (DM) generation relies on direct couplings to the visible sector that are strong enough to induce a thermal equilibirum in the early universe. In recent years however the ever increasing experimental constraints have cast doubts whether this truly is the mechanism behind the observed DM component \cite{Arcadi:2017kky} and other tangential ideas have gained in popularity.

In many ways a minimal set up is postulating a singlet scalar field, which has been proposed as a possible DM candidate already in \cite{McDonald:1993ex,Matos:1998vk}. Very weak interactions between the visible and dark sectors are sufficient for a working DM phenomenology, as realised in the FIMP framework \cite{Giudice:1999yt,Hall:2009bx}. The dark matter field could even reside in a completely decoupled sector connected to the visible one only through gravity.
%A natural approach is hence one where dark matter is given by a scalar field residing in a completely decoupled sector connected to the visible one only through gravity. 
Particle production, even if the sector is decoupled, is unavoidable for a light scalar field in the early universe as already discussed in Refs. \cite{Ford:1986sy,Starobinsky:1986fx,Starobinsky:1994bd}. The vacuum fluctuations of a light decoupled scalar have been shown to be a viable explanation of dark matter in Refs.~\cite{Kuzmin:1998kk} and \cite{Peebles:1999fz}, with or without self-interactions, respectively. Related set-ups are fuzzy dark matter \cite{Hu:2000ke} and axion dark matter, e.g. Ref.~\cite{Marsh:2010wq} and reviewed in \cite{Marsh:2015xka}, which have been extensively studied in literature. Super heavy Wimpzillas are yet another  viable DM candidate that can be produced gravitationally in the early universe~\cite{ Kolb:1998ki,Chung:1998zb,Peebles:1999fz,Kolb:2017jvz,Li:2019ves}. Dark matter can also be sourced by graviton mediated scatterings \cite{Garny:2015sjg,Garny:2017kha}.

For quantum fields in curved spaces a non-minimal coupling to curvature generically emerges~\cite{Chernikov:1968zm,Callan:1970ze,Tagirov:1972vv}. Such couplings have non-trivial implications for inflation \cite{Kaiser:2013sna,Schutz:2013fua,Takahashi:2020car,vandeVis:2020qcp} and importantly, as was shown in Refs.~\cite{Figueroa:2016dsc,Dimopoulos:2018wfg,Nakama:2018gll,Opferkuch:2019zbd,Kamada:2019ewe,Bettoni:2018utf,Bettoni:2019dcw},  non-minimal curvature terms can source matter generation. As expected, dark matter can also be produced via a non-minimal coupling, and the production can take place during inflation \cite{Alonso-Alvarez:2018tus,AlonsoAlvarez:2019cgw}, or during the reheating phase \cite{Markkanen:2015xuw,Fairbairn:2018bsw,Ema:2016hlw,Ema:2018ucl,Velazquez:2019mpj}.

Many gravitational DM generation mechanisms result in an inherently isocurvature component (see \cite{Byrnes:2006fr} for example), which is heavily constrained by observations \cite{Akrami:2018odb}. Correct estimate of isocurvature is thus crucial \cite{Chung:2004nh,Markkanen:2018gcw,Padilla:2019fju}. In some cases the gravitationally generated DM is however strictly adiabatic  \cite{Markkanen:2015xuw,Fairbairn:2018bsw}, although large couplings may be needed for its efficient production. 

In this work we investigate a gravitational DM set-up where the field is heavy enough not to be disturbed during inflation, but the energy density is generated via spontaneous symmetry breaking induced by the non-minimal curvature coupling. What we will show is that the mechanism is efficient even for ${\cal O}(1)$ values of the (negative) non-minimal coupling and, importantly, that the perturbations are completely adiabatic. We provide both analytical estimates and perform a full numerical study of the produced dark matter abundance, using the quadratic and Starobinsky models \cite{Linde:1983gd,Starobinsky:1980te} as templates for the inflaton ptential. We use of the $(+,+,+)$ sign conventions of Ref.~\cite{Misner:1974qy}. 

{\bf Note added:} While we were preparing this manuscript, a closely related work studying a similar mechanism with analytical approximations appeared \cite{Babichev:2020xeg}. However, in \cite{Babichev:2020xeg} it was assumed that the singlet bare mass is larger than the Hubble rate at the onset of singlet oscillations. We do not make this assumption, in fact we concentrate on the opposite limit, which explains why our quantitative results differ from \cite{Babichev:2020xeg}.

\section{The setup and primordial dynamics}
\label{sec:model}
 
We investigate a scalar singlet $\chi$ with the $Z_{2}$ symmetric action  
\beq
\label{chiaction}
S_{\chi}=-\int d^4x\,\sqrt{|g|}\left(\frac{1}{2}(\nabla\chi)^2+\frac{1}{2}m^2\chi^2+\frac{\xi}{2}R\chi^2+\frac{\lambda}{4}\chi^4\right)~.
\eeq
Here $R$ is the scalar curvature and the non-minimal coupling $\xi R\chi^2$ is necessary to obtain a renormalisable theory in a curved spacetime (note that our sign convention for $\xi$ in eq. (\ref{chiaction}) is the opposite of e.g. \cite{Bezrukov:2007ep}). We assume the energy density of $\chi$ is negligible 
during inflation, reheating and early radiation domination so that it can be treated as a test field in this epoch. 

%We assume $\chi$ is a spectator field during inflation and reheating, meaning that its energy density is negligible $\rho(\chi) \ll 3 H^3 M_{\rm P}^2$ and the field can be quantized as a test field in a classical background spacetime. In this setup the $\beta$ functions for the spectator couplings are given by \cite{...}
%\baq
%\beta_{\xi}&=&\frac{1}{16\pi^2} 6\lambda\left(\xi-\frac{1}{6}\right)~,\\
%\beta_{m^2}&=&\frac{1}{16\pi^2} 6\lambda m^2~,\\
%\beta_{\lambda}&=&\frac{1}{16\pi^2} 18\lambda^2~.\\
%\eaq 
%It should be noted that  the non-minimal term $\xi R\chi^2$ is necessarily to renormalise the theory in a curved classical spacetime background. In particular, since $\xi =0$ is not a fixed point, the non-minimal coupling $\xi$ cannot be made to vanish over a range of scales unless the self-coupling $\lambda$ also vanishes.

%\section{Dynamics during inflation}
%\label{sec:ssb}

The curvature scalar can be written as $R  = 12 H^2 (1-\epsilon_{\rm H}/2)$ where $\epsilon_{\rm H} =-\dot{H}/H^2$.  
During inflation $\epsilon_{\rm H} \ll 1$ and $R \simeq 12 H^2 $ gives a nearly constant contribution to the effective mass of $\chi$. For 
$\xi < 0$ and $m^2 < 12 |\xi| H^2$
the tree-level effective potential for $\chi$ develops $Z_{2}$ symmetry breaking minima at $\chi = \pm \chi_*$ where
\beq
\label{chistar}
\chi_* = \left(\frac{-12 \xi H^2-m^2}{\lambda}\right)^{1/2} ~,
\eeq
and the equality holds to leading order in $\epsilon_{\rm H}$.  Our test field assumption remains valid at the minima  provided that the (negative) potential energy satisfies  $|V(\chi_*)|\ll  3H_{*}^2M_{\rm P}^2$.  A sufficient condition for this is  
${|\xi|}/{\sqrt{\lambda}} \ll {M_{\rm P}}/{H}$, and since the bound on the tensor-to-scalar ratio \cite{Akrami:2018odb} implies  $M_{\rm P}/{H} \gtrsim 10^{5}$ during the observable period of inflation, we find that the test field assumption holds in a very large range of coupling values. 

The dynamics around the minimum depends on the effective mass. If $V''(\chi_*)/H^2 < 9/4$ the field is light and  quantum fluctuations grow in the infrared regime \cite{Starobinsky:1994bd}. Reliable analysis of the dynamics in this case requires non-pertrubative resummation techniques such as the stochastic formalism \cite{Starobinsky:1994bd} which is beyond the scope of the current work. Here 
we restrict ourselves to the opposite limit, $V''(\chi_*)/H^2 > 9/4$, which corresponds to  
\beq
\label{massivecond}
|\xi| > \frac{3}{32} + \frac{1}{12}\frac{m^2}{H^2} ~. 
\eeq
In this case the field is massive around the minimum and there is no infrared amplification of quantum fluctuations. Starting from an initial condition in the vicinity of $\pm\chi_{*}$ the local field field value rapidly relaxes to the minimum and tracks the classical time-evolution $\pm \chi_*(t)$ with a small lag due to the effective friction $3H\dot{\chi}$. The global structure of the field space depends on the epoch before inflation. If $R=0$ (e.g. due to conformal symmetry), the $Z_{2}$ symmetry is spontaneously broken at the onset of inflation $t_{\rm in}$ and the size of $\pm\chi_{*}$ domains scales proportional to $(a_{\rm in}H_{\rm in})^{-1}$ in comoving units. This is exponentially larger than the currently observable universe, $(a_0 H_0)/(a_{\rm in} H_{\rm in}) \sim  e^{N_{\rm tot}-N_{\rm obs}}$, if the total number of efolds $N_{\rm tot}$ is larger than $N_{\rm obs}\sim 60$. On the other hand, if the symmetry was not restored before inflation, the global structure depends on whatever is the mechanism that sets the initial conditions for the universe. Either way, here we simply assume that the comoving size of $\pm \chi_*$ domains at the onset of inflation exceeds the size of the currently observable universe so that we do not need to account for possible domain walls in our analysis. 

Since the location of the minimum (\ref{chistar}) is determined by the Hubble rate, its fluctuations during inflation will be directly inherited by $\chi$. This leads to formation of adiabatic perturbations in the singlet energy density $\rho_{\chi} = -\lambda \chi_{*}^4(H)/4$.  In particular, for slow roll inflation with a single inflaton $\phi$ we get 
\beq
\frac{\delta\rho_\chi}{\dot{\rho}_{\chi}} = \frac{\rho_{\chi}'(H)\delta H}{\rho_{\chi}'(H)\dot{H}} = 
-\frac{\delta H}{\epsilon_{\rm H} H^2}=\frac{\delta\rho_\phi}{\dot{\rho}_{\phi}}~, 
\eeq    
explicitly showing the absence of isocurvature perturbations between the inflaton sector and $\chi$. Therefore, any relic of the $\chi$ sector will constitute an adiabatic dark matter component. We stress that the assumption of massive limit (\ref{massivecond}) is crucial here. In the opposite massless limit, which we do not consider here, there would be  uncorrelated superhorizon perturbations of $\chi$ on top of the adiabatic perturbations which would give rise to isocurvature perturbations heavily constrained by the data \cite{Akrami:2018odb}.

\section{Analytical estimate of the dark matter abundance}

\label{sec:analytical}

After the end of inflation the universe reheats and evolves towards radiation domination. In the radiation dominated phase, $R=0$ (up to quantum corrections, which we neglect) and the symmetry breaking minimum is removed. However, depending on the evolution of $R(t)$ over reheating, the field $\chi$ may not just smoothly relax from the inflationary value (\ref{chistar}) to the symmetric minimum $\chi =0$ but it can end up oscillating around it. In fact, this is always the case in the concrete setups that we study here. Since the singlet cannot decay, the oscillations continue indefinitely. As the amplitude gets redshifted, the potential eventually becomes dominated by the quadratic term and we are left with a dark relic component whose energy density scales like  non-relativistic matter. In this section, we make some qualitative remarks on the evolution and present a rough analytical estimate for dark matter relic abundance based on simplistic approximations. Results of the full numerical analysis are presented in the next section.

As mentioned above, during inflation $\chi$ tracks the time-evolving minimum (\ref{chistar}) with a small lag. More concretely, writing $\chi(t) = \chi_*(t)+\tilde{\chi}(t)$ (choosing the positive minimum $+\chi_*$ for definiteness), and expanding the equation of motion to first order in $\tilde{\chi}\ll \chi_*$, we get 
\beq
\label{Deltachieom}
\ddot{\tilde{\chi}} + 3 H \dot{\tilde{\chi}} + V''(\chi_{*}) \tilde{\chi} = -\ddot{\chi_{*}} - 3 H \dot{\chi_{*}}~.
\eeq
Concentrating for simplicity on the limit $m^2 \ll |\xi| H^2 $, equation (\ref{chistar}) yields $\ddot{\chi_{*}} - 3 H \dot{\chi_{*}} \simeq 3\epsilon_{\rm H}  H^2 \chi_*$ to leading order in slow roll. As long as the time dependence of $\bar{\chi}(t)$ is weak, we can drop the first two terms on the left hand side of eq. (\ref{Deltachieom}) to get   
\beq
\frac{\chi-\chi_*}{\chi_*}\sim \epsilon_{\rm H} \frac{3H^2}{V''(\chi_*)} \ll 1~.
\eeq
This approximative expression is valid during inflation and in the massive limit (\ref{massivecond}) where $V''(\chi_*)/H^2 > 9/4$ for $\epsilon_{\rm H} \ll 1$. The lag $\chi(t)-\chi_*(t)$ is the smaller the larger the mass $V''(\chi_*)$ because increasing the mass makes cosmic friction less efficient. 

\begin{figure}[!ht]
\begin{center}
\includegraphics[width=0.9\textwidth]{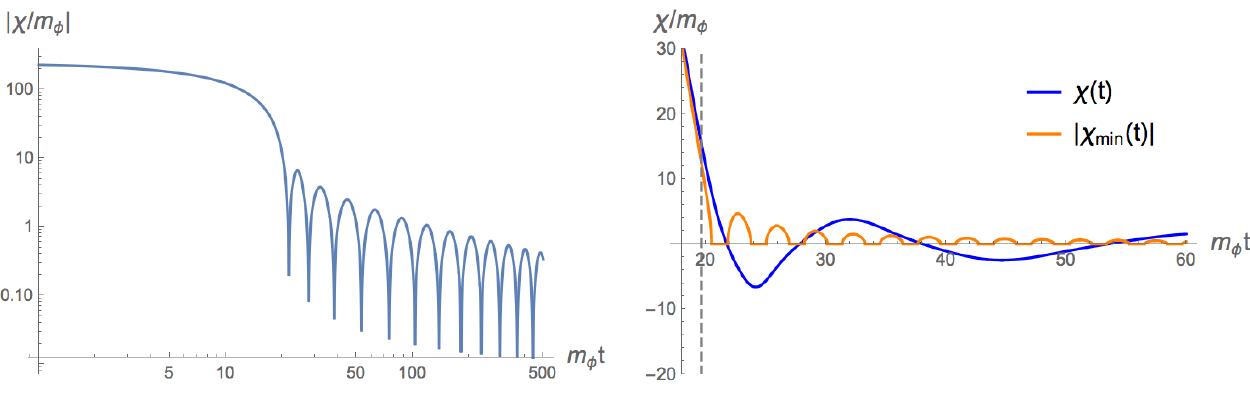}
\caption{Left panel: $\chi(t)$ computed numerically for $\xi =-1, m = 1\;{\rm MeV}, \lambda = 0.01 $, and a quadratic inflaton potential $m_{\phi} = 1.53\times 10^{13}\; {\rm GeV}$ and a constant inflaton decay rate into radiation $\Gamma = 1.00\times 10^{11}\; {\rm GeV}$ (see next section for further details). Right panel: the same in linear scale, zoomed in to the onset of oscillations and showing also the evolution of the minimum $V'(\chi_{\rm min}) = 0$. The dashed line marks the end of inflation $\epsilon_{\rm H} = 1$. }
\label{fig:chidynamics}
\end{center}
\end{figure}

At the end of inflation $\epsilon_{\rm H}$ starts to grow decreasing the ratio $V''(\chi_*)/H^2 = 24 |\xi| (1-\epsilon_{\rm H}/2)$ (note that here we use the full expression $R=12 H^2 (1-\epsilon_{\rm H}/2)$ unlike in eq. (\ref{chistar}) where the slow roll suppressed part can be suspended), and causing $\chi(t)$ to gradually deviate from $\chi_*(t)$.  In this work we assume inflation is followed by a reheating epoch driven by an oscillating inflaton field $\phi$. The curvature scalar $R =M_{\rm P}^{-2}(4V_{\phi}(\phi)-\dot{\phi}^2)$ then becomes a decreasing function which oscillates from positive to negative values (assuming the inflaton potential vanishes in the vacuum). The minima of the resulting time dependent effective potential of $\chi$ are located at $\chi = \pm ((-\xi R -m^2)/\lambda)^{1/2}$ when $R > -m^2/\xi$ (recall that $\xi <0$) and at $\chi = 0$ when $R < -m^2/\xi$. This leads to oscillatory dynamics for $\chi(t)$, as illustrated in figure \ref{fig:chidynamics}.   

\begin{figure}[!ht]
\begin{center}
\includegraphics[width=0.9\textwidth]{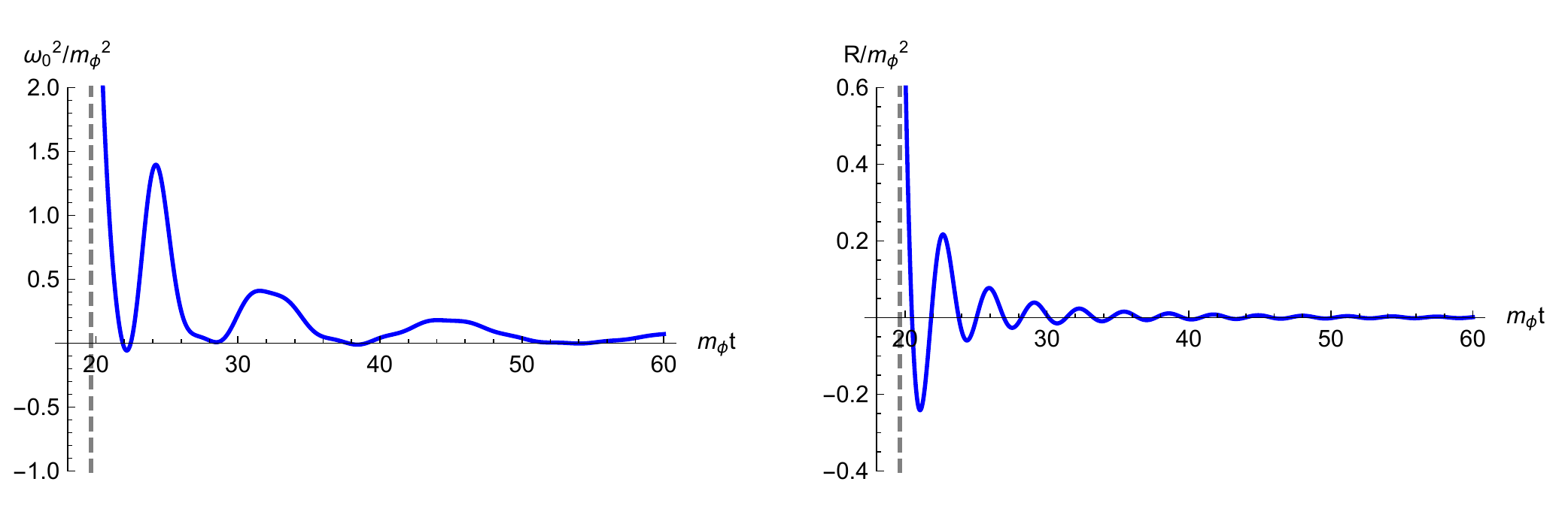}
\caption{ Left panel: effective mass squared $\omega^2_0 = -H^2/4-\dot{H}/2+m^2+3\lambda\chi^2+(\xi-1/6) R$ for the $k=0$ mode of the fluctuation  around the classical solution, $v_{k} \equiv a^{3/2} \delta \chi_{k}$, which obeys $\ddot{v}_k+\omega^2_k v_k=0$ \cite{Fairbairn:2018bsw}. The irregular features are not numerical error but caused by the evolution of $R(t)$. Same parameter values as in Fig. 1. Right panel: evolution of the curvature scalar in the same setup.}
\label{fig:meff}
\end{center}
\end{figure}

One may ask if the oscillation of $R(t)$ from positive to negative values could induce tachyonic particle production, or spinodal decomposition, which would lead to exponential growth of the two point function $\langle\chi^2\rangle$ and potentially significant deviations from the classical evolution of the one point function. This is known to happen in related setups with non-minimally coupled scalars \cite{Bassett:1997az,Tsujikawa:1999jh,Markkanen:2015xuw,Fairbairn:2018bsw} (see \cite{Dufaux:2006ee} for tachyonic preheating scenarios with non-gravitaional operators and \cite{Bernal:2018hjm} for a DM application). In our setup with $\xi <0$, the effective mass squared of fluctuations around the classical solution can be negative only in the vicinity of $\chi = 0$ and only when the broken minimum $|\chi_*| = ((-\xi R -m^2)/\lambda)^{1/2}$ exists, i.e. when $R > -m^2/\xi$. At the beginning of oscillations $|\chi(t)| >|\chi_*(t)|$, and the broken minimum gets lifted $|\chi_*(t)|\rightarrow 0$ before the field $\chi(t)$ reaches zero for the first time. This heavily suppresses the tachyonic regime where the effective mass squared is negative. As the oscillations proceed, the curvature scalar $R$ decreases faster than the effective mass contribution proportional to $\lambda\chi^2$, and the tachyonic regime gets almost completely removed, see figure \ref{fig:meff} showing numerically computed classical evolution of the effective mass squared of fluctuations and $R$ for a choice of parameters. 
The situation might change for $|\xi|\gg 1$, when on the one hand the tachyonic regimes may become larger (see the discussion at the end of Section \ref{sec:abundance}), and on the other the exponential growth of $\langle\chi^2\rangle $ with the exponent proportional to $\sqrt{|\xi]} m_{\phi} \Delta t $ ($m_{\phi}$ being the inflaton mass) might compensate for the narrowness of the tachyonic window $\Delta t \ll m_{\phi}^{-1}$. However, in this work we will mainly concentrate on the regime $|\xi]\sim {\cal O}(1)$ (or smaller) where we expect that the tachyonic particle production does not lead to significant deviations from the classical analysis we perform.  

When the reheating completes, the universe enters the radiation dominated phase where $R =0$ and $\chi$ continues to oscillate around zero in the potential $V(\chi) = m^2\chi^2/2 + \lambda \chi^4/4$. To get the most na\"{i}ve analytical order of magnitude estimate for the envelope of oscillations at the end of reheating, $\bar{\chi}_{\rm reh}$, we may use 
\beq
\label{chirehapp}
\bar{\chi}_{\rm reh} \sim \sqrt{\frac{-6 \xi}{\lambda}} H_{\rm end} \frac{a_{\rm end}}{a_{\rm reh}}~.
\eeq
This effectively corresponds to setting $R$ to zero immediately at the end of inflaton $t_{\rm end}$ such that $\chi$ starts to oscillate in a quartic potential $V(\chi) = \lambda \chi^4/4$ with the initial amplitude given by $\chi = (-12\xi H_{\rm end}^2 (1-\epsilon_{\rm H}/2)/\lambda)^{1/2}$ where $\epsilon_{\rm H}= 1$. This is clearly not the correct picture and, as discussed in more detail in the next section, the final estimate for the relic abundance obtained using (\ref{chirehapp}) differs from the actual numerical result by a $\xi$ dependent factor typically ${\cal O}(10)$. However, this level of accuracy suffices for the qualitative discussion in this section. Note also that although the mass term $m^2\chi^2$ could a priori be comparable to $\lambda\chi^4$ during reheating, this case is phenomenologically excluded as $\chi$ would start to behave as non-relativistic matter already in the very early universe leading to massive overproduction of dark matter. 

After the end of reheating, the envelope scales as $\bar{\chi}\propto a^{-1}$ until the potential becomes dominated by the quadratic bare mass term and the scaling changes to $\bar{\chi}\propto a^{-3/2}$. The transition time  $t_{\rm tr}$ between the two regimes is well approximated by  $\lambda \bar{\chi}_{\rm tr}^4/4  = m^2\bar{\chi}_{\rm tr}^2/2$. By matching the two scaling solutions at $t_{\rm tr}$, the envelope in the asymptotic regime $t \gg t_{\rm tr}$ is given by   
\beq
\label{chiscaling}
\bar{\chi}(t) =  \left(\frac{\lambda}{2}\right)^{1/4}\frac{\bar{\chi}_{\rm reh}^{3/2}}{m^{1/2}}
\left(\frac{a_{\rm reh}}{a(t)}\right)^{3/2}~. 
\eeq
The energy density of the oscillating field in this regime is simply  $\rho_{\chi} = 1/2 m^2 \bar{\chi}^2 \propto a^{-3}$ and field therefore constitutes a dark relic component behaving like non-relativistic matter as far as the energy budget is concerned. Using eq. (\ref{chirehapp}) for $\bar{\chi}_{\rm reh}$ and approximating the equation of state during reheating by a constant $w$ we the obtain a rough estimate for the dark matter relic density today 
\beq
\label{rhochiappr}
\Omega_{\chi} h^2 \sim 0.2|\xi|^{3/2}\frac{ m}{\rm MeV}\left(\frac{\lambda}{0.01}\right)^{-1}
\left(\frac{H_{\rm end}}{10^{13}{\rm GeV}}\right)^{3/2} \left(\frac{g_{*\rm reh}}{100}\right)^{-1/4}
\left(\frac{H_{\rm reh}}{H_{\rm end}}\right)^{\frac{1-3w}{2+2w}}~.
\eeq 
Here $t_{\rm reh}$ denotes the time when reheating is completed and the radiation dominated epoch starts, $g_{*\rm reh}$ denotes number of relativistic degrees of freedom at $t_{\rm reh}$, we have approximated $g_{\rm *}\sim g_{*\rm s}$ and used $g_{*{\rm s},0} = 3.909$, $T_0 = 2.725 {\rm K}$ and $H_0/h \equiv 100{\rm km}/{\rm s}/{\rm Mpc}$. The deviation of the estimate (\ref{rhochiappr}) from the numerical result is discussed in the next section.

Note that to get viable cosmology, the transition to the quadratic regime should take place before the matter radiation equality, $T_{\rm tr} >  0.8 {\rm eV}$ where $T$ is the plasma temperature. Using that $\bar{\chi}_{\rm tr} = m (2/\lambda)^{1/2} = \bar{\chi}_{\rm reh}(a_{\rm reh}/a_{\rm tr})$, and substituting (\ref{chirehapp}) for $\bar{\chi}_{\rm reh}$, this translates to the meV scale bound  
\beq
\label{matradeqbound}
\frac {m}{0.001 {\rm eV}} \gtrsim \sqrt{|\xi|}\left(\frac{H_{\rm end}}{10^{13}{\rm GeV}}\right)^{1/2}\left(\frac{H_{\rm end}}{H_{\rm reh}}\right)^{\frac{-1+3w}{6+6w}}\frac{g_{*\rm tr}^{1/3}}{g_{*\rm reh}^{1/12}}~.
\eeq
Note also that for $m \gtrsim {\rm meV}$, the non-zero curvature term $\xi R\chi^2$ induced again during matter domination $R=3H^2$ and late time dark energy domination $R\simeq 12 H^2$, is completely negligible compared to the bare mass term as $R < 3H_{\rm eq}^2 \lll m^2$. Hence there will be no new symmetry breaking and we can neglect the curvature term in the analysis. 

\section{Numerical analysis}

\label{sec:abundance}

In this section we compute the dark matter relic abundance for the model (\ref{chiaction}) by numerically solving the set of classical equations of motion given by
\baq
\label{eomset}
\ddot{\phi}+3H\dot{\phi} + \Gamma \dot{\phi} + V_{\phi}'(\phi)  &=&0\\
\ddot{\chi}+3H\dot{\chi} + \lambda\chi^3 +m^2\chi + \xi R \chi &=&0\\\nonumber
\dot{\rho}_{\rm r} + 4 H \rho_{\rm r} &=& \Gamma \dot{\phi}^2\\\nonumber 
3 H^2 M_{\rm P}^2 &=& \frac{1}{2}\dot{\phi}^2 + V_{\phi}(\phi) + \rho_{\rm r}\\\nonumber
R&=&M_{\rm pl}^{-2}(4V_{\phi} (\phi)  -\dot{\phi}^2)\nonumber~.
\eaq
As templates for the inflaton sector we consider the quadratic inflation and $R^2$ model specified respectively by\footnote{Strictly speaking the second model in eq.~(\ref{mods}) is not exactly the $R^2$ model \cite{Starobinsky:1979ty,Kehagias:2013mya} as there non-renormalizable couplings between the inflaton and all other sectors are generated.  If included, one would anyway expect such terms to be sub-dominant after inflation as they are Planck suppressed.}
\begin{align}
V_{\phi}=\frac{1}{2}m_{\phi}^2\phi^2~,\qquad  {\rm and} \qquad V_{\phi}=\Lambda_{\phi}^4(1-e^{-\sqrt{{2}/{3}}{\phi}/{M_{\rm P}}})^2~.\label{mods}
\end{align}
In both cases we use a constant rate $\Gamma$ to model the inflaton decay into radiation.

We set slow roll initial conditions for the inflaton sector and choose $m_{\phi}$ and $\Gamma$  ($\Lambda_{\phi}$ and $\Gamma$) such that the amplitude of perturbations equals the observed value $P_{\cal R} = 2.1 \times 10^{-9}$  at the pivot scale $k = 0.05 {\rm Mpc}^{-1}$ \cite{Aghanim:2018eyx}. For the singlet we set the initial conditions $\chi(t_{\rm in}) = \chi_*(t_{\rm in}), \dot{\chi}(t_{\rm in}) = 0$ from which the system rapidly relaxes to the tracking solution $\chi(t)=\chi_*(t) + \tilde{\chi}(t)$ with $\tilde{\chi}\ll \chi_*$. We choose the initial time $N(t_{\rm in}) =70$ efolds before the end of inflation which is more than enough to ensure that $\chi(t)$ has relaxed to the tracking solution well before the end of inflation. 

In principle we could numerically evolve the set of equations (\ref{eomset}) until the regime where the singlet potential is dominated by the bare mass term $m^2\chi^2$ and $\rho_{\chi} \propto a^3$. However, as noted above, the  phenomenologically interesting cases which avoid extreme overproduction of dark matter correspond to $m \ll H_{\rm reh}$. One would thus need to evolve the rapidly oscillating system over a very long time after the end of reheating $t_{\rm reh}$ which in practice becomes too heavy numerically. To avoid this, we choose instead to evolve the system until a reference time $t_{1}$ well after the end of reheating and thereafter use the analytical scaling law (\ref{chiscaling}) with $\bar{\chi}_{\rm reh}$ replaced by the numerically evaluated input value $\bar{\chi}_1$. We have checked that this agrees with the full numerical result to better than 5\%  accuracy in the case of (phenomenologically uninteresting) large values of $m$ for which the numerical solution can be run into the quadratic regime. More importantly, as the approximation concerns late time evolution only, it gives the same systematic error for all values of $\xi$ and all reheating histories. We choose $t_{1}$ as the moment when $H(t_1) = 0.01 \Gamma$ which is well in the radiation domination and well before $t_{\rm tr}$ for the entire parameter range shown in our results. We have also checked that the results do not depend on the precise choice of $t_1$. Using that $g_{*{\rm s}}a^3 T^3$ is constant, we can then finally express the singlet relic abundance today $t_0$ in the form 
\beq
\label{abundance}
\Omega_{\chi} h^2  =  \frac{\sqrt{\lambda}}{2\sqrt{2}}\frac{m\bar{\chi}_1^3}{3 (H_0/h)^2 M_{\rm P}^2}\frac{g_{*{\rm s,0}}T_0^3}{g_{*{\rm s},1}T_1^3}~,
\eeq
where $\bar{\chi}_1$ and $T_1$ are determined numerically, and $g_{*{\rm s},0} = 3.909$, $T_0 = 2.725 {\rm K}$ and $H_0/h = 100{\rm km}/{\rm s}/{\rm Mpc}$. 

\begin{figure}[!ht]
\begin{center}
\includegraphics[width=1.0\textwidth]{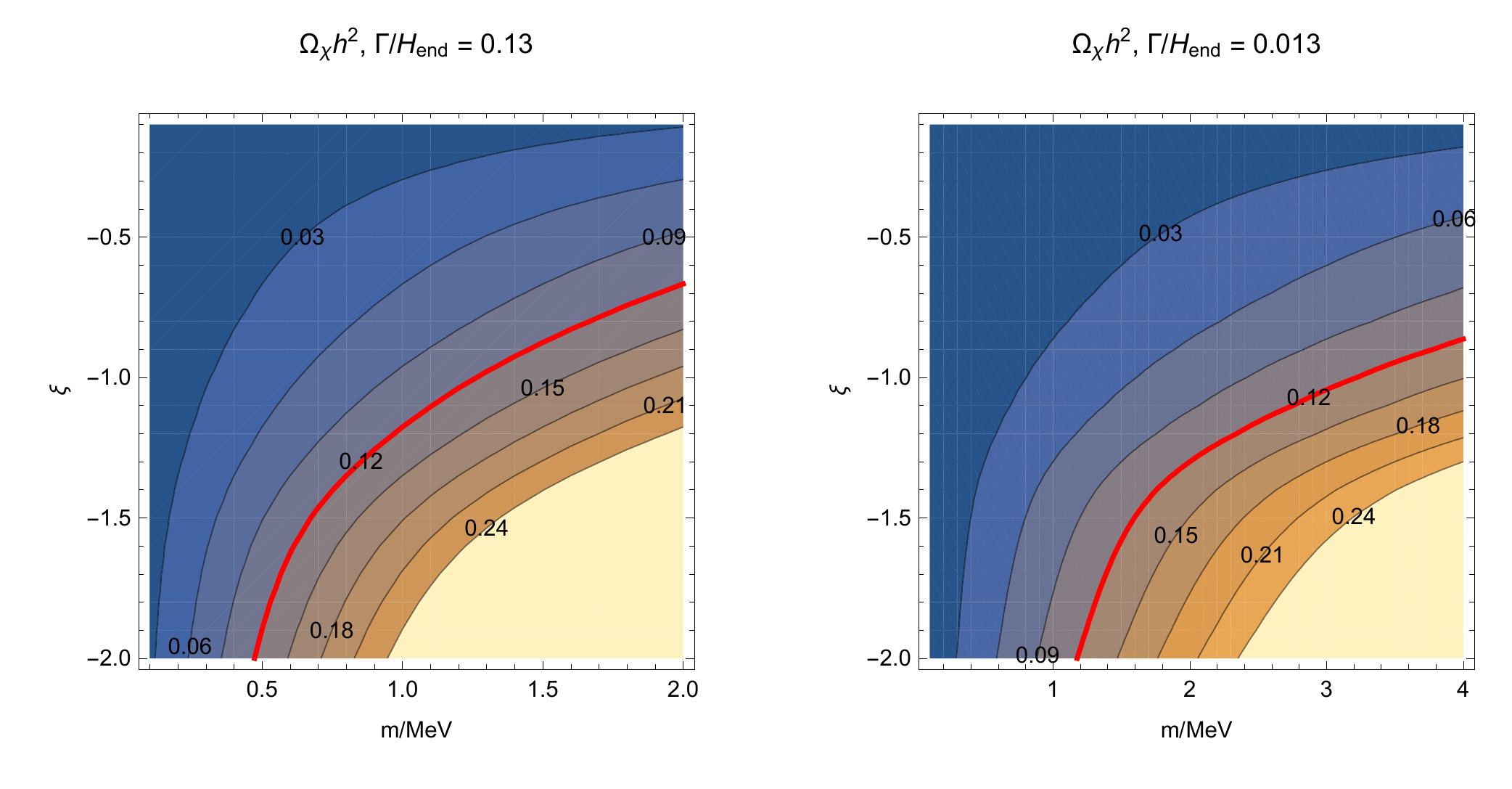}
\caption{Dark matter relic abundance as function of the relic mass $m_{\chi}$ and its non-minimal coupling $\xi$ for the self-coupling value $\lambda = 0.01$ computed using the quadratic inflaton potential. The red line marks the contour where the dark matter abundance equals the measured value $\Omega_{\rm c}h^2 =0.12$. In the left panel $m_{\phi} = 1.52\times 10^{13}\; {\rm GeV}$ and $\Gamma = 1.00\times 10^{12}\; {\rm GeV}$. In the right panel $m_{\phi} = 1.53\times 10^{13}\; {\rm GeV}$ and $\Gamma = 1.00\times 10^{11}\; {\rm GeV}$.}
\label{fig:phi2plots}
\end{center}
\end{figure}

\begin{figure}[!ht]
\begin{center}
\includegraphics[width=1.0\textwidth]{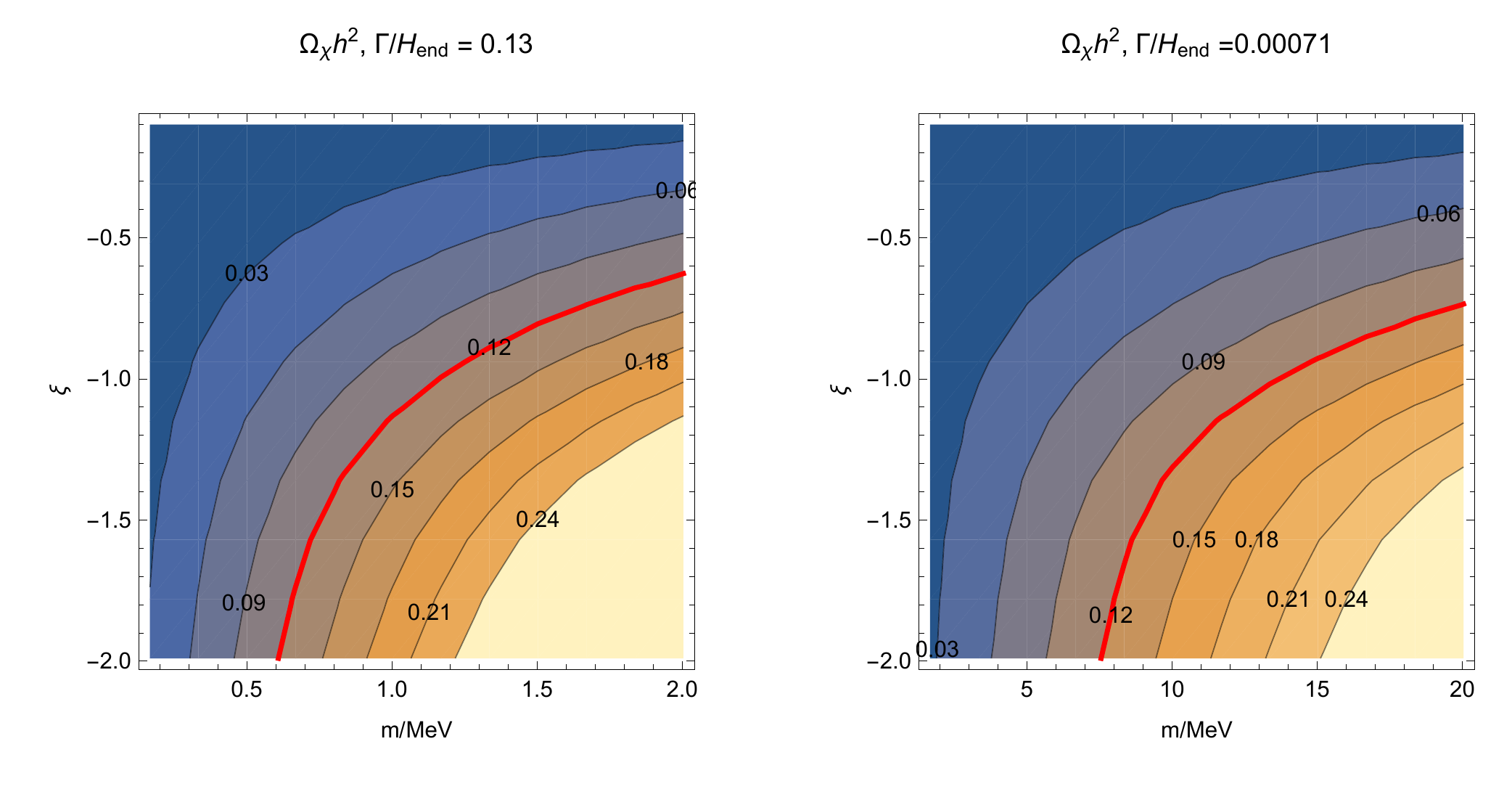}
\caption{Dark matter relic abundance as function of the relic mass $m_{\chi}$ and its non-minimal coupling $\xi$ for the self-coupling value $\lambda = 0.01$ computed using the $R^2$ inflaton potential. The red line marks the contour where the dark matter abundance equals the measured value $\Omega_{\rm c}h^2 =0.12$.  In the left panel $\Lambda_{\phi} = 7.79\times 10^{15}\; {\rm GeV}$ and $\Gamma = 9.00\times 10^{11}\; {\rm GeV}$. In the right panel $\Lambda_{\phi} = 7.81\times 10^{15}\; {\rm GeV}$ and $\Gamma = 5.00\times 10^{9}\; {\rm GeV}$.}
\label{fig:r2plots}
\end{center}
\end{figure}

The main results of this work are figures \ref{fig:phi2plots} and \ref{fig:r2plots} which show the numerically computed relic abundance as function the relic mass $m$ and the non-minimal coupling $\xi$. 
Fig. \ref{fig:phi2plots} shows the results for the quadratic inflaton potential and two choices of $\Gamma$, one corresponding to instant reheating with $H = \Gamma$ reached during the first inflaton oscillation, and the other  corresponding to more delayed reheating with 12 inflaton oscillations from the end of inflation until the moment $H = \Gamma$. Fig. \ref{fig:r2plots} shows the results for the $R^2$ inflaton potential and for two choices of $\Gamma$. The first again corresponds to instant reheating during the first inflaton oscillation (close to the estimated duration of reheating in Higgs inflation \cite{Ema:2016dny,Sfakianakis:2018lzf,DeCross:2015uza,DeCross:2016fdz,DeCross:2016cbs}, see also \cite{Repond:2016sol,Bezrukov:2008ut,GarciaBellido:2008ab}). In the other example, we have chosen $\Gamma$ to yield  ${\cal O}(220)$ inflaton oscillations from the end of inflation until $H =\Gamma$. The inflaton parameters $m_{\phi}$ and $\Lambda_{\phi}$, respectively, are set to give the observed amplitude of perturbations for the chosen value of $\Gamma$. In both cases, we find that relic masses in the MeV scale yield the observed dark matter abundance $\Omega_{\rm c}h^2 =0.12$ \cite{Aghanim:2018eyx} for $|\xi|\sim {\cal O}(1)$, and the self coupling value $\lambda = 0.01$ chosen in the figures. 

As can be seen in the figures, lowering the reheating temperature by decreasing $\Gamma$ by an order of magnitude decreases the abundance by roughly a factor of three when all other parameters are kept fixed. This is well in agreement with the analytical expression (\ref{rhochiappr}) (where $w=0$ for both the quadratic and $R^2$ potential when the inflaton oscillates around the minimum). Also the $m$ dependence seen in the figures agrees well with the scaling given by eq. (\ref{rhochiappr}). This is expected since the mass term has a negligible effect during reheating and this part of eq. (\ref{rhochiappr}) should therefore be unaffected by the approximations made in eq. (\ref{chirehapp}). We have also checked that the scaling of eq. (\ref{rhochiappr}) as function of $\lambda$ and $H_{\rm end}$ agrees well with the numerical result. This can be understood by the following argument. By rescaling the field and time as $\tilde{\chi} = \chi/\chi_{\rm in}$, $\tilde{t} = t R_{\rm in}^{1/2}$, and neglecting the bare mass term, it can be seen that the initial values enter the equation of motion of the spectator $\tilde{\chi}$ only through $R/R_{\rm in}$ and the combination $\lambda \chi_{\rm in}^2/R_{\rm in}$ from which both $\lambda$ and $H_{\rm in}$ cancel out using eq. (\ref{chistar}) for $\chi_{\rm in}$ and $R_{\rm in} = 12 H_{\rm in}^2$. This explains why the the simple expression (\ref{rhochiappr}) describes well the dependence on $\lambda$ and $H_{\rm end}$. 

On the other hand, as expected, the numerical results show a significantly more complicated dependence on $\xi$ than the estimate eq. (\ref{rhochiappr}), obtained neglecting effects of the oscillating mass term $\xi R$ during reheating. 
\begin{figure}[!ht]
\begin{center}
\includegraphics[width=0.9\textwidth]{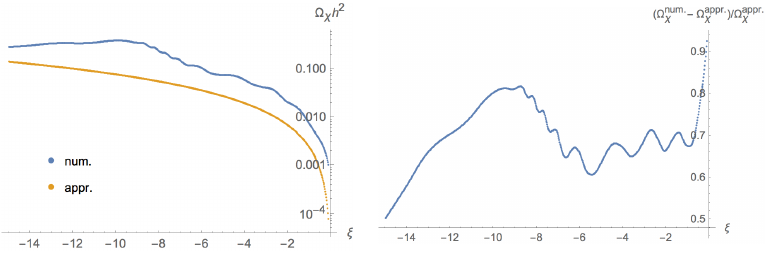}
\caption{The numerically computed relic abundance as function of $\xi$ vs. the analytical estimate (\ref{rhochiappr}) (left panel) and their relative difference (right panel). Computed using $m = 0.2 {\rm MeV},\lambda = 0.01$ and quadratic inflaton potential with $m_{\phi} = 1.53\times 10^{13}\; {\rm GeV}$ and $\Gamma = 1.00\times 10^{11}\; {\rm GeV}$.}
\label{fig:xiplot}
\end{center}
\end{figure}
Figure \ref{fig:xiplot} shows the $\xi$ dependence of the full numerical result and eq. (\ref{rhochiappr}), and their relative difference, for a choice of $m, \lambda$ and the quadratic inflaton potential. There are two main effects which lead to the rather complicated behaviour of the numerical result. First, increasing $|\xi|$ makes $\chi$ more massive initially, causing it to track the moving minimum longer and effectively start oscillating with a smaller amplitude. For the setup shown in Fig. 5, this is the dominant effect for $|\xi|\lesssim 1$. Second, the formation of temporary broken minima when $R$ oscillates to positive values may either slow down or accelerate the redshifting of the amplitude of $\chi$, depending on the time-evolving phase difference between $R(t)$ and $\chi(t)$. 

For larger couplings $|\xi|$, the curvature term may also become large enough to trap $\chi(t)$ oscillating around one of the broken minima for one or more cycles of oscillation, in Fig. \ref{fig:xiplot} this happens for $|\xi| \gtrsim 10$. In this regime, the windows where tachyonic particle production can take place also start to get longer which may call for a more careful study beyond the classical approach which however is outside the scope of this work.

\section{Conclusions}

\label{sec:conclusions}

In this work we have shown that adiabatic dark matter can arise from a gravitationally induced symmetry breaking during inflation. As an  example, we have studied a $Z_{2}$ symmetric scalar singlet model including the non-minimal curvature coupling $\xi R\chi^2$ required for renormalizability.  For negative values of $\xi$, the $Z_{2}$ symmetry gets broken during inflation, when $R \simeq 12 H^2$, and restored when the universe reheats and becomes radiation dominated with $R=0$. This leads to dynamics where the singlet initially sits in the broken minimum $\chi_*$ until the end of inflation and then starts to oscillate around $\chi = 0$, at least if reheating starts with the usual inflaton oscillations. If the singlet has no couplings to other matter fields, it oscillates indefinitely with a redshifting amplitude, and asymptotically probes only the bare mass part of its potential $m^2\chi^2/2$. The oscillating field  will then  constitute a dark relic component with the energy density scaling as non-relativistic matter.

We have performed a detailed numerical study of the scenario, computing the relic abundance for two different inflaton potentials, and provided  approximative analytical expressions to build qualitative understanding of the results. We have concentrated in the regime $\xi \lesssim -3/32 $ where the singlet is effectively heavy during inflation and perturbations of the dark matter component are adiabatic in accordance with observations.  Our main results are summarised in figures \ref{fig:phi2plots} and \ref{fig:r2plots} which show the dark matter abundance as function of the singlet mass $m$ and the non-minimal coupling $\xi$, computed for reheating dynamics driven by  quadratic and $R^2$ inflaton potentials, respectively. We find that the observed relic abundance $\Omega_{\rm c}h^2 =0.12$ is obtained for singlet masses in the MeV scale and non-minimal coupling values $|\xi| = {\cal O}(1)$. This is for the singlet four point coupling $\lambda = {\cal O}( 0.01)$ and the abundance scales proportional to $\lambda^{-1}$. 

Our result demonstrates that, interestingly, adiabatic dark matter can arise from a sector  coupled to other fields only through gravity, masses below the electroweak scale and perturbatively small coupling values. Since such gravitationally produced dark relics could escape all direct detection constraints, it would be extremely important to understand their impacts on the formation of large scale structures in the universe. 

\acknowledgments

L.L. is supported by the Magnus Ehrnrooth Foundation  and T.M. by the Estonian Research Council (Mobilitas Plus MOBJD323).

\bibliography{dm}

\providecommand{\href}[2]{#2}\begingroup\raggedright\begin{thebibliography}{10}

\bibitem{Arcadi:2017kky}
G.~Arcadi, M.~Dutra, P.~Ghosh, M.~Lindner, Y.~Mambrini, M.~Pierre et~al.,
  \emph{{The waning of the WIMP? A review of models, searches, and
  constraints}},
  \href{http://dx.doi.org/10.1140/epjc/s10052-018-5662-y}{\emph{Eur. Phys. J.
  C} {\bf 78} (2018) 203}, [\href{http://arxiv.org/abs/1703.07364}{{\tt
  1703.07364}}].

\bibitem{McDonald:1993ex}
J.~McDonald, \emph{{Gauge singlet scalars as cold dark matter}},
  \href{http://dx.doi.org/10.1103/PhysRevD.50.3637}{\emph{Phys. Rev. D} {\bf
  50} (1994) 3637--3649}, [\href{http://arxiv.org/abs/hep-ph/0702143}{{\tt
  hep-ph/0702143}}].

\bibitem{Matos:1998vk}
T.~Matos and F.~S. Guzman, \emph{{Scalar fields as dark matter in spiral
  galaxies}}, \href{http://dx.doi.org/10.1088/0264-9381/17/1/102}{\emph{Class.
  Quant. Grav.} {\bf 17} (2000) L9--L16},
  [\href{http://arxiv.org/abs/gr-qc/9810028}{{\tt gr-qc/9810028}}].

\bibitem{Giudice:1999yt}
G.~Giudice, I.~Tkachev and A.~Riotto, \emph{{Nonthermal production of dangerous
  relics in the early universe}},
  \href{http://dx.doi.org/10.1088/1126-6708/1999/08/009}{\emph{JHEP} {\bf 08}
  (1999) 009}, [\href{http://arxiv.org/abs/hep-ph/9907510}{{\tt
  hep-ph/9907510}}].

\bibitem{Hall:2009bx}
L.~J. Hall, K.~Jedamzik, J.~March-Russell and S.~M. West, \emph{{Freeze-In
  Production of FIMP Dark Matter}},
  \href{http://dx.doi.org/10.1007/JHEP03(2010)080}{\emph{JHEP} {\bf 03} (2010)
  080}, [\href{http://arxiv.org/abs/0911.1120}{{\tt 0911.1120}}].

\bibitem{Ford:1986sy}
L.~Ford, \emph{{Gravitational Particle Creation and Inflation}},
  \href{http://dx.doi.org/10.1103/PhysRevD.35.2955}{\emph{Phys. Rev. D} {\bf
  35} (1987) 2955}.

\bibitem{Starobinsky:1986fx}
A.~A. Starobinsky, \emph{{STOCHASTIC DE SITTER (INFLATIONARY) STAGE IN THE
  EARLY UNIVERSE}},
  \href{http://dx.doi.org/10.1007/3-540-16452-9\_6}{\emph{Lect. Notes Phys.}
  {\bf 246} (1986) 107--126}.

\bibitem{Starobinsky:1994bd}
A.~A. Starobinsky and J.~Yokoyama, \emph{Equilibrium state of a selfinteracting
  scalar field in the de sitter background},
  \href{http://dx.doi.org/10.1103/PhysRevD.50.6357}{\emph{Phys.Rev.D} {\bf 50}
  (1994) 6357--6368}, [\href{http://arxiv.org/abs/astro-ph/9407016}{{\tt
  astro-ph/9407016}}].

\bibitem{Kuzmin:1998kk}
V.~Kuzmin and I.~Tkachev, \emph{{Matter creation via vacuum fluctuations in the
  early universe and observed ultrahigh-energy cosmic ray events}},
  \href{http://dx.doi.org/10.1103/PhysRevD.59.123006}{\emph{Phys. Rev. D} {\bf
  59} (1999) 123006}, [\href{http://arxiv.org/abs/hep-ph/9809547}{{\tt
  hep-ph/9809547}}].

\bibitem{Peebles:1999fz}
P.~Peebles and A.~Vilenkin, \emph{{Noninteracting dark matter}},
  \href{http://dx.doi.org/10.1103/PhysRevD.60.103506}{\emph{Phys. Rev. D} {\bf
  60} (1999) 103506}, [\href{http://arxiv.org/abs/astro-ph/9904396}{{\tt
  astro-ph/9904396}}].

\bibitem{Hu:2000ke}
W.~Hu, R.~Barkana and A.~Gruzinov, \emph{{Cold and fuzzy dark matter}},
  \href{http://dx.doi.org/10.1103/PhysRevLett.85.1158}{\emph{Phys. Rev. Lett.}
  {\bf 85} (2000) 1158--1161},
  [\href{http://arxiv.org/abs/astro-ph/0003365}{{\tt astro-ph/0003365}}].

\bibitem{Marsh:2010wq}
D.~J. Marsh and P.~G. Ferreira, \emph{{Ultra-Light Scalar Fields and the Growth
  of Structure in the Universe}},
  \href{http://dx.doi.org/10.1103/PhysRevD.82.103528}{\emph{Phys. Rev. D} {\bf
  82} (2010) 103528}, [\href{http://arxiv.org/abs/1009.3501}{{\tt 1009.3501}}].

\bibitem{Marsh:2015xka}
D.~J.~E. Marsh, \emph{{Axion Cosmology}},
  \href{http://dx.doi.org/10.1016/j.physrep.2016.06.005}{\emph{Phys. Rept.}
  {\bf 643} (2016) 1--79}, [\href{http://arxiv.org/abs/1510.07633}{{\tt
  1510.07633}}].

\bibitem{Kolb:1998ki}
E.~W. Kolb, D.~J.~H. Chung and A.~Riotto, \emph{{WIMPzillas!}},
  \href{http://dx.doi.org/10.1063/1.59655}{\emph{AIP Conf. Proc.} {\bf 484}
  (1999) 91--105}, [\href{http://arxiv.org/abs/hep-ph/9810361}{{\tt
  hep-ph/9810361}}].

\bibitem{Chung:1998zb}
D.~J.~H. Chung, E.~W. Kolb and A.~Riotto, \emph{{Superheavy dark matter}},
  \href{http://dx.doi.org/10.1103/PhysRevD.59.023501}{\emph{Phys. Rev.} {\bf
  D59} (1999) 023501}, [\href{http://arxiv.org/abs/hep-ph/9802238}{{\tt
  hep-ph/9802238}}].

\bibitem{Kolb:2017jvz}
E.~W. Kolb and A.~J. Long, \emph{{Superheavy dark matter through Higgs portal
  operators}}, \href{http://dx.doi.org/10.1103/PhysRevD.96.103540}{\emph{Phys.
  Rev. D} {\bf 96} (2017) 103540}, [\href{http://arxiv.org/abs/1708.04293}{{\tt
  1708.04293}}].

\bibitem{Li:2019ves}
L.~Li, T.~Nakama, C.~M. Sou, Y.~Wang and S.~Zhou, \emph{{Gravitational
  Production of Superheavy Dark Matter and Associated Cosmological
  Signatures}}, \href{http://dx.doi.org/10.1007/JHEP07(2019)067}{\emph{JHEP}
  {\bf 07} (2019) 067}, [\href{http://arxiv.org/abs/1903.08842}{{\tt
  1903.08842}}].

\bibitem{Garny:2015sjg}
M.~Garny, M.~Sandora and M.~S. Sloth, \emph{{Planckian Interacting Massive
  Particles as Dark Matter}},
  \href{http://dx.doi.org/10.1103/PhysRevLett.116.101302}{\emph{Phys. Rev.
  Lett.} {\bf 116} (2016) 101302}, [\href{http://arxiv.org/abs/1511.03278}{{\tt
  1511.03278}}].

\bibitem{Garny:2017kha}
M.~Garny, A.~Palessandro, M.~Sandora and M.~S. Sloth, \emph{{Theory and
  Phenomenology of Planckian Interacting Massive Particles as Dark Matter}},
  \href{http://dx.doi.org/10.1088/1475-7516/2018/02/027}{\emph{JCAP} {\bf 02}
  (2018) 027}, [\href{http://arxiv.org/abs/1709.09688}{{\tt 1709.09688}}].

\bibitem{Chernikov:1968zm}
N.~Chernikov and E.~Tagirov, \emph{{Quantum theory of scalar fields in de
  Sitter space-time}}, {\emph{Ann. Inst. H. Poincare Phys. Theor. A} {\bf 9}
  (1968) 109}.

\bibitem{Callan:1970ze}
J.~Callan, Curtis~G., S.~R. Coleman and R.~Jackiw, \emph{{A New improved energy
  - momentum tensor}},
  \href{http://dx.doi.org/10.1016/0003-4916(70)90394-5}{\emph{Annals Phys.}
  {\bf 59} (1970) 42--73}.

\bibitem{Tagirov:1972vv}
E.~Tagirov, \emph{{Consequences of field quantization in de Sitter type
  cosmological models}},
  \href{http://dx.doi.org/10.1016/0003-4916(73)90047-X}{\emph{Annals Phys.}
  {\bf 76} (1973) 561--579}.

\bibitem{Kaiser:2013sna}
D.~I. Kaiser and E.~I. Sfakianakis, \emph{{Multifield Inflation after Planck:
  The Case for Nonminimal Couplings}},
  \href{http://dx.doi.org/10.1103/PhysRevLett.112.011302}{\emph{Phys. Rev.
  Lett.} {\bf 112} (2014) 011302}, [\href{http://arxiv.org/abs/1304.0363}{{\tt
  1304.0363}}].

\bibitem{Schutz:2013fua}
K.~Schutz, E.~I. Sfakianakis and D.~I. Kaiser, \emph{{Multifield Inflation
  after Planck: Isocurvature Modes from Nonminimal Couplings}},
  \href{http://dx.doi.org/10.1103/PhysRevD.89.064044}{\emph{Phys. Rev. D} {\bf
  89} (2014) 064044}, [\href{http://arxiv.org/abs/1310.8285}{{\tt 1310.8285}}].

\bibitem{Takahashi:2020car}
T.~Takahashi, T.~Tenkanen and S.~Yokoyama, \emph{{Violation of slow-roll in
  non-minimal inflation}},  \href{http://arxiv.org/abs/2003.10203}{{\tt
  2003.10203}}.

\bibitem{vandeVis:2020qcp}
J.~van~de Vis, R.~Nguyen, E.~I. Sfakianakis, J.~T. Gibiln and D.~I. Kaiser,
  \emph{{Time-Scales for Nonlinear Processes in Preheating after Multifield
  Inflation with Nonminimal Couplings}},
  \href{http://arxiv.org/abs/2005.00433}{{\tt 2005.00433}}.

\bibitem{Figueroa:2016dsc}
D.~G. Figueroa and C.~T. Byrnes, \emph{{The Standard Model Higgs as the origin
  of the hot Big Bang}},
  \href{http://dx.doi.org/10.1016/j.physletb.2017.01.059}{\emph{Phys. Lett. B}
  {\bf 767} (2017) 272--277}, [\href{http://arxiv.org/abs/1604.03905}{{\tt
  1604.03905}}].

\bibitem{Dimopoulos:2018wfg}
K.~Dimopoulos and T.~Markkanen, \emph{{Non-minimal gravitational reheating
  during kination}},
  \href{http://dx.doi.org/10.1088/1475-7516/2018/06/021}{\emph{JCAP} {\bf 06}
  (2018) 021}, [\href{http://arxiv.org/abs/1803.07399}{{\tt 1803.07399}}].

\bibitem{Nakama:2018gll}
T.~Nakama and J.~Yokoyama, \emph{{Reheating through the Higgs amplified by
  spinodal instabilities and gravitational creation of gravitons}},
  \href{http://dx.doi.org/10.1093/ptep/ptz014}{\emph{PTEP} {\bf 2019} (2019)
  033E02}, [\href{http://arxiv.org/abs/1803.07111}{{\tt 1803.07111}}].

\bibitem{Opferkuch:2019zbd}
T.~Opferkuch, P.~Schwaller and B.~A. Stefanek, \emph{{Ricci Reheating}},
  \href{http://dx.doi.org/10.1088/1475-7516/2019/07/016}{\emph{JCAP} {\bf 07}
  (2019) 016}, [\href{http://arxiv.org/abs/1905.06823}{{\tt 1905.06823}}].

\bibitem{Kamada:2019ewe}
K.~Kamada, J.~Kume, Y.~Yamada and J.~Yokoyama, \emph{{Gravitational
  leptogenesis with kination and gravitational reheating}},
  \href{http://dx.doi.org/10.1088/1475-7516/2020/01/016}{\emph{JCAP} {\bf 01}
  (2020) 016}, [\href{http://arxiv.org/abs/1911.02657}{{\tt 1911.02657}}].

\bibitem{Bettoni:2018utf}
D.~Bettoni and J.~Rubio, \emph{{Quintessential Affleck-Dine baryogenesis with
  non-minimal couplings}},
  \href{http://dx.doi.org/10.1016/j.physletb.2018.07.046}{\emph{Phys. Lett. B}
  {\bf 784} (2018) 122--129}, [\href{http://arxiv.org/abs/1805.02669}{{\tt
  1805.02669}}].

\bibitem{Bettoni:2019dcw}
D.~Bettoni and J.~Rubio, \emph{{Hubble-induced phase transitions: Walls are not
  forever}}, \href{http://dx.doi.org/10.1088/1475-7516/2020/01/002}{\emph{JCAP}
  {\bf 01} (2020) 002}, [\href{http://arxiv.org/abs/1911.03484}{{\tt
  1911.03484}}].

\bibitem{Alonso-Alvarez:2018tus}
G.~Alonso-Álvarez and J.~Jaeckel, \emph{{Lightish but clumpy: scalar dark
  matter from inflationary fluctuations}},
  \href{http://dx.doi.org/10.1088/1475-7516/2018/10/022}{\emph{JCAP} {\bf 10}
  (2018) 022}, [\href{http://arxiv.org/abs/1807.09785}{{\tt 1807.09785}}].

\bibitem{AlonsoAlvarez:2019cgw}
G.~Alonso-Álvarez, J.~Jaeckel and T.~Hugle, \emph{{Misalignment \& Co.:
  (Pseudo-)scalar and vector dark matter with curvature couplings}},
  \href{http://dx.doi.org/10.1088/1475-7516/2020/02/014}{\emph{JCAP} {\bf 02}
  (2020) 014}, [\href{http://arxiv.org/abs/1905.09836}{{\tt 1905.09836}}].

\bibitem{Markkanen:2015xuw}
T.~Markkanen and S.~Nurmi, \emph{{Dark matter from gravitational particle
  production at reheating}},
  \href{http://dx.doi.org/10.1088/1475-7516/2017/02/008}{\emph{JCAP} {\bf 02}
  (2017) 008}, [\href{http://arxiv.org/abs/1512.07288}{{\tt 1512.07288}}].

\bibitem{Fairbairn:2018bsw}
M.~Fairbairn, K.~Kainulainen, T.~Markkanen and S.~Nurmi, \emph{{Despicable Dark
  Relics: generated by gravity with unconstrained masses}},
  \href{http://dx.doi.org/10.1088/1475-7516/2019/04/005}{\emph{JCAP} {\bf 04}
  (2019) 005}, [\href{http://arxiv.org/abs/1808.08236}{{\tt 1808.08236}}].

\bibitem{Ema:2016hlw}
Y.~Ema, R.~Jinno, K.~Mukaida and K.~Nakayama, \emph{{Gravitational particle
  production in oscillating backgrounds and its cosmological implications}},
  \href{http://dx.doi.org/10.1103/PhysRevD.94.063517}{\emph{Phys. Rev. D} {\bf
  94} (2016) 063517}, [\href{http://arxiv.org/abs/1604.08898}{{\tt
  1604.08898}}].

\bibitem{Ema:2018ucl}
Y.~Ema, K.~Nakayama and Y.~Tang, \emph{{Production of Purely Gravitational Dark
  Matter}}, \href{http://dx.doi.org/10.1007/JHEP09(2018)135}{\emph{JHEP} {\bf
  09} (2018) 135}, [\href{http://arxiv.org/abs/1804.07471}{{\tt 1804.07471}}].

\bibitem{Velazquez:2019mpj}
J.~A. Cembranos, L.~J. Garay and J.~M. Sánchez~Velázquez,
  \emph{{Gravitational production of scalar dark matter}},
  \href{http://arxiv.org/abs/1910.13937}{{\tt 1910.13937}}.

\bibitem{Byrnes:2006fr}
C.~T. Byrnes and D.~Wands, \emph{{Curvature and isocurvature perturbations from
  two-field inflation in a slow-roll expansion}},
  \href{http://dx.doi.org/10.1103/PhysRevD.74.043529}{\emph{Phys. Rev. D} {\bf
  74} (2006) 043529}, [\href{http://arxiv.org/abs/astro-ph/0605679}{{\tt
  astro-ph/0605679}}].

\bibitem{Akrami:2018odb}
{\scshape Planck} collaboration, Y.~Akrami et~al., \emph{{Planck 2018 results.
  X. Constraints on inflation}},  \href{http://arxiv.org/abs/1807.06211}{{\tt
  1807.06211}}.

\bibitem{Chung:2004nh}
D.~J. Chung, E.~W. Kolb, A.~Riotto and L.~Senatore, \emph{{Isocurvature
  constraints on gravitationally produced superheavy dark matter}},
  \href{http://dx.doi.org/10.1103/PhysRevD.72.023511}{\emph{Phys. Rev. D} {\bf
  72} (2005) 023511}, [\href{http://arxiv.org/abs/astro-ph/0411468}{{\tt
  astro-ph/0411468}}].

\bibitem{Markkanen:2018gcw}
T.~Markkanen, A.~Rajantie and T.~Tenkanen, \emph{Spectator dark matter},
  \href{http://dx.doi.org/10.1103/PhysRevD.98.123532}{\emph{Phys.Rev.D} {\bf
  98} (2018) 123532}, [\href{http://arxiv.org/abs/1811.02586}{{\tt
  1811.02586}}].

\bibitem{Padilla:2019fju}
L.~E. Padilla, J.~A. Vázquez, T.~Matos and G.~Germán, \emph{{Scalar Field
  Dark Matter Spectator During Inflation: The Effect of Self-interaction}},
  \href{http://dx.doi.org/10.1088/1475-7516/2019/05/056}{\emph{JCAP} {\bf 05}
  (2019) 056}, [\href{http://arxiv.org/abs/1901.00947}{{\tt 1901.00947}}].

\bibitem{Linde:1983gd}
A.~D. Linde, \emph{{Chaotic Inflation}},
  \href{http://dx.doi.org/10.1016/0370-2693(83)90837-7}{\emph{Phys. Lett. B}
  {\bf 129} (1983) 177--181}.

\bibitem{Starobinsky:1980te}
A.~A. Starobinsky, \emph{{A New Type of Isotropic Cosmological Models Without
  Singularity}},
  \href{http://dx.doi.org/10.1016/0370-2693(80)90670-X}{\emph{Adv. Ser.
  Astrophys. Cosmol.} {\bf 3} (1987) 130--133}.

\bibitem{Misner:1974qy}
C.~W. Misner, K.~Thorne and J.~Wheeler, \emph{{Gravitation}}.
\newblock W. H. Freeman, San Francisco, 1973.

\bibitem{Babichev:2020xeg}
E.~Babichev, D.~Gorbunov and S.~Ramazanov, \emph{{Gravitational misalignment
  mechanism of Dark Matter production}},
  \href{http://arxiv.org/abs/2004.03410}{{\tt 2004.03410}}.

\bibitem{Bezrukov:2007ep}
F.~L. Bezrukov and M.~Shaposhnikov, \emph{{The Standard Model Higgs boson as
  the inflaton}},
  \href{http://dx.doi.org/10.1016/j.physletb.2007.11.072}{\emph{Phys. Lett. B}
  {\bf 659} (2008) 703--706}, [\href{http://arxiv.org/abs/0710.3755}{{\tt
  0710.3755}}].

\bibitem{Bassett:1997az}
B.~A. Bassett and S.~Liberati, \emph{{Geometric reheating after inflation}},
  \href{http://dx.doi.org/10.1103/PhysRevD.60.049902}{\emph{Phys. Rev. D} {\bf
  58} (1998) 021302}, [\href{http://arxiv.org/abs/hep-ph/9709417}{{\tt
  hep-ph/9709417}}].

\bibitem{Tsujikawa:1999jh}
S.~Tsujikawa, K.-i. Maeda and T.~Torii, \emph{{Resonant particle production
  with nonminimally coupled scalar fields in preheating after inflation}},
  \href{http://dx.doi.org/10.1103/PhysRevD.60.063515}{\emph{Phys. Rev. D} {\bf
  60} (1999) 063515}, [\href{http://arxiv.org/abs/hep-ph/9901306}{{\tt
  hep-ph/9901306}}].

\bibitem{Dufaux:2006ee}
J.~F. Dufaux, G.~N. Felder, L.~Kofman, M.~Peloso and D.~Podolsky,
  \emph{{Preheating with trilinear interactions: Tachyonic resonance}},
  \href{http://dx.doi.org/10.1088/1475-7516/2006/07/006}{\emph{JCAP} {\bf 07}
  (2006) 006}, [\href{http://arxiv.org/abs/hep-ph/0602144}{{\tt
  hep-ph/0602144}}].

\bibitem{Bernal:2018hjm}
N.~Bernal, A.~Chatterjee and A.~Paul, \emph{{Non-thermal production of Dark
  Matter after Inflation}},
  \href{http://dx.doi.org/10.1088/1475-7516/2018/12/020}{\emph{JCAP} {\bf 12}
  (2018) 020}, [\href{http://arxiv.org/abs/1809.02338}{{\tt 1809.02338}}].

\bibitem{Starobinsky:1979ty}
A.~A. Starobinsky, \emph{{Spectrum of relict gravitational radiation and the
  early state of the universe}}, {\emph{JETP Lett.} {\bf 30} (1979) 682--685}.

\bibitem{Kehagias:2013mya}
A.~Kehagias, A.~Moradinezhad~Dizgah and A.~Riotto, \emph{{Remarks on the
  Starobinsky model of inflation and its descendants}},
  \href{http://dx.doi.org/10.1103/PhysRevD.89.043527}{\emph{Phys. Rev. D} {\bf
  89} (2014) 043527}, [\href{http://arxiv.org/abs/1312.1155}{{\tt 1312.1155}}].

\bibitem{Aghanim:2018eyx}
{\scshape Planck} collaboration, N.~Aghanim et~al., \emph{{Planck 2018 results.
  VI. Cosmological parameters}},  \href{http://arxiv.org/abs/1807.06209}{{\tt
  1807.06209}}.

\bibitem{Ema:2016dny}
Y.~Ema, R.~Jinno, K.~Mukaida and K.~Nakayama, \emph{{Violent Preheating in
  Inflation with Nonminimal Coupling}},
  \href{http://dx.doi.org/10.1088/1475-7516/2017/02/045}{\emph{JCAP} {\bf 02}
  (2017) 045}, [\href{http://arxiv.org/abs/1609.05209}{{\tt 1609.05209}}].

\bibitem{Sfakianakis:2018lzf}
E.~I. Sfakianakis and J.~van~de Vis, \emph{{Preheating after Higgs Inflation:
  Self-Resonance and Gauge boson production}},
  \href{http://dx.doi.org/10.1103/PhysRevD.99.083519}{\emph{Phys. Rev. D} {\bf
  99} (2019) 083519}, [\href{http://arxiv.org/abs/1810.01304}{{\tt
  1810.01304}}].

\bibitem{DeCross:2015uza}
M.~P. DeCross, D.~I. Kaiser, A.~Prabhu, C.~Prescod-Weinstein and E.~I.
  Sfakianakis, \emph{{Preheating after Multifield Inflation with Nonminimal
  Couplings, I: Covariant Formalism and Attractor Behavior}},
  \href{http://dx.doi.org/10.1103/PhysRevD.97.023526}{\emph{Phys. Rev. D} {\bf
  97} (2018) 023526}, [\href{http://arxiv.org/abs/1510.08553}{{\tt
  1510.08553}}].

\bibitem{DeCross:2016fdz}
M.~P. DeCross, D.~I. Kaiser, A.~Prabhu, C.~Prescod-Weinstein and E.~I.
  Sfakianakis, \emph{{Preheating after multifield inflation with nonminimal
  couplings, II: Resonance Structure}},
  \href{http://dx.doi.org/10.1103/PhysRevD.97.023527}{\emph{Phys. Rev. D} {\bf
  97} (2018) 023527}, [\href{http://arxiv.org/abs/1610.08868}{{\tt
  1610.08868}}].

\bibitem{DeCross:2016cbs}
M.~P. DeCross, D.~I. Kaiser, A.~Prabhu, C.~Prescod-Weinstein and E.~I.
  Sfakianakis, \emph{{Preheating after multifield inflation with nonminimal
  couplings, III: Dynamical spacetime results}},
  \href{http://dx.doi.org/10.1103/PhysRevD.97.023528}{\emph{Phys. Rev. D} {\bf
  97} (2018) 023528}, [\href{http://arxiv.org/abs/1610.08916}{{\tt
  1610.08916}}].

\bibitem{Repond:2016sol}
J.~Repond and J.~Rubio, \emph{{Combined Preheating on the lattice with
  applications to Higgs inflation}},
  \href{http://dx.doi.org/10.1088/1475-7516/2016/07/043}{\emph{JCAP} {\bf 07}
  (2016) 043}, [\href{http://arxiv.org/abs/1604.08238}{{\tt 1604.08238}}].

\bibitem{Bezrukov:2008ut}
F.~Bezrukov, D.~Gorbunov and M.~Shaposhnikov, \emph{{On initial conditions for
  the Hot Big Bang}},
  \href{http://dx.doi.org/10.1088/1475-7516/2009/06/029}{\emph{JCAP} {\bf 06}
  (2009) 029}, [\href{http://arxiv.org/abs/0812.3622}{{\tt 0812.3622}}].

\bibitem{GarciaBellido:2008ab}
J.~Garcia-Bellido, D.~G. Figueroa and J.~Rubio, \emph{{Preheating in the
  Standard Model with the Higgs-Inflaton coupled to gravity}},
  \href{http://dx.doi.org/10.1103/PhysRevD.79.063531}{\emph{Phys. Rev. D} {\bf
  79} (2009) 063531}, [\href{http://arxiv.org/abs/0812.4624}{{\tt 0812.4624}}].

\end{thebibliography}\endgroup

\end{document}